\documentclass[a4paper,12pt]{article}
\usepackage[T2A]{fontenc}
\usepackage[cp1251]{inputenc}
\usepackage[dvips]{graphicx}
\usepackage{amssymb,amsmath}
\numberwithin{equation}{section}

\begin{document}

\title{ON THE INTEGRABLE GENERALIZATION OF THE 1D  TODA LATTICE}
\author{P.Yu.Tsyba, K.R.Esmakhanova,   G.N.Nugmanova, R.Myrzakulov\footnote{The corresponding author. Email: cnlpmyra1954@yahoo.com} \\ \textit{Eurasian International Center for Theoretical Physics,} \\ \textit{Eurasian National University, Astana, 010008, Kazakhstan}}

\date{}

\maketitle
\begin{abstract}
A generalized Toda Lattice equation is considered. The associated
linear problem (Lax representation) is found. For simple case
$N=3$ the $\tau$-function Hirota form is presented that allows to
construct  an exast solutions of the equations of the 1DGTL. The
corresponding hierarchy and its relations with the nonlinear
Schrodinger equation and Hersenberg ferromagnetic equation are
discussed.
\end{abstract}
\vspace{2cm} \sloppy
%\tableofcontents
%\newpage
\section{Introduction} %\label{intro}

The 1D Toda lattice (1DTL)
\begin{equation} \ddot
q_n=e^{q_{n-1}-q_{n}}-e^{q_{n}-q_{n+1}}
\end{equation} is one of most
important integrable equations which plays an important role in
mathematics and physics. In physics, the equations (1.1) describe
an interacting $N$ particles, each with mass $m_{n}=1$, arranged
along a line at positions $q_{1}, q_{2}, ... , q_{N}$. Between
each pair of adjacent particles, there is a force whose magnitude
depends exponentially on the distance between them. The 1DTL was
discovered by Morikasu Toda in 1967 \cite{toda}. Using the
computer experiments, in \cite{ford} was suggested that the 1DTL
is integrable. In \cite{flaschka1}, \cite{henon}, \cite{manakov}
the integrability of the 1DTL is proved.  Note that this equation
is a discrete approximation of the KdV equation
\begin{equation}
u_t+6uu_x+u_{xxx}=0. \label{Hamiltonian}
\end{equation}

The aim of this Letter is to construct the some integrable
generalization of the equation (1.1), that can be linked to both
linear and quadratic compatible Poissonbrackets for the usual
1DTL.
\section{Background on 1DTL}\label{Background on 1DTL}

In this section we present some  known fundamental informations
for the 1DTL and fix some notations.  Let $p_{n}$ denotes the
momentum of the $n$th particle. Then the total energy of the
system is the Hamiltonian
\begin{equation}
H=\frac{1}{2}\sum^{N}_{n=1}p_{n}^{2}+\sum_{n=1}^{N-1}e^{q_{n}-q_{n+1}}.
 \label{Hamiltonian} \end{equation}

So the system (1.1) can be written as
\begin{equation}\label{Hamiltonian DEs}
\begin{array}{lcl}
\dot q_{n}&=&\{H,q_{n}\}=\frac{\partial H}{\partial p_{n}},\\
\dot p_{n}&=&\{H,p_{n}\}=-\frac{\partial H}{\partial q_{n}}.
\end{array}\end{equation}
Except the original form (1.1), there are exist its the other
various equivalent forms. Some of them as follows.
\\
i) \begin{equation}\label{Hamiltonian DEs}
\begin{array}{lcl}
\dot \alpha_{n}&=&\alpha_{n}(\beta_{n}-\beta_{n+1}),\\
\dot \beta_{n}&=&\alpha_{n-1}-\alpha_{n}.
\end{array}\end{equation}
 ii) \begin{equation}
\begin{array}{lcl}
\dot a_{n}&=&a_{n}(b_{n+1}-b_{n}),\\
\dot b_{n}&=&2(\alpha_{n}^2-\alpha_{n-1}^2).
\end{array}\end{equation}
iii)\begin{equation}\label{Hamiltonian DEs}
\begin{array}{lcl}
[D_{t}^{2}-4\sinh^{2}(\frac{D_{n}}{2}))]f_{n}\circ f_{n}=0.
\end{array}\end{equation}
iv) \begin{equation}\label{Hamiltonian DEs}
\ddot\tau_{n}\tau_{n}-\dot\tau^{2}_{n}=\tau_{n+1}\tau_{n-1}\end{equation}
and so on. Above  $D_{t}, \ D_{n}$ are the well known Hirota
bilinear operators and  $\tau_{n}$ is so-called  $\tau$-function
which play a key role of the theory of integrable systems. Note
that a new and initial "physical" ($q_{n}, p_{n}$) dependent
variables are related   as
\begin{equation}\label{Hamiltonian DEs}
\begin{array}{lcl}
 \alpha_{n}&=&e^{q_{n}-q_{n+1}},\quad\quad \quad \ \ \beta_{n}=p_{n},\\
 a_{n}&=&\frac{1}{2}e^{\frac{1}{2}(q_{n}-q_{n+1})},\quad\quad
 b_{n}=-\frac{1}{2}p_{n},\\
 \ddot\ln{ f_{n}}&=& (\ln f_{n})_{tt}=e^{q_{n-1}-q_{n}},\\
 q_{n}&=&\ln{\frac{\tau_{n-1}}{\tau_{n}}}.
\end{array}\end{equation}
There are at least two possible Lax representations  for the 1DTL,
one of order 2$\times$2, another one of order N$\times$N (see,
e.i. \cite{Su1}). The 2$\times$2 Lax pair is defined as
\begin{equation}L_n=
\begin{pmatrix}  p_{n}+\lambda &  e^{q_n} \cr
                   -e^{-q_n} &0     \end{pmatrix}, \quad M_n=
\begin{pmatrix} 0 &  -e^{q_n} \cr
                   e^{-q_n} &\lambda      \end{pmatrix}.
\end{equation}
The associated linear problem is
\begin{equation}\label{Hamiltonian DEs}
\begin{array}{lcl}
 \Psi_{n+1}&=&L_{n}\Psi_{n},\\
 \Psi_{nt}&=&M_{n}\Psi_{n}.
\end{array}\end{equation} The compatibility condition of these equations
 \begin{equation}\label{Hamiltonian DEs}
L_{nt}+L_{n}M_{n}-M_{n+1}L_{n}=0\end{equation} gives the equation
(1.1). The equivalent Lax representation is given by  the
N$\times$N Lax pair
\begin{equation}
L_n=\sum_{\nu=1}^{N}L^{\nu}X_{\nu}, \quad
M_n=\sum_{\nu=1}^{N}M^{\nu}X_{\nu}, \label{}
\end{equation}
where $[X_{\mu},X_{\nu}]=C_{\mu\nu}^{\lambda}X_{\lambda}$. Then
the TL equation is obtained from the Lax equation
\begin{equation}\label{Hamiltonian DEs}
\dot L=[L,M]. \end{equation} There  are exist a so-called r-matrix
representation for the Poisson brackets $\{L^{\mu},L^{\nu}\}$
between the matrix elements of $L$. It has the form
\begin{equation}\label{Hamiltonian DEs}
\{L\otimes_{,}L\}=[r, L\otimes I+I\otimes L]=[r,L_{1}+L_{2}]
\end{equation}  or
\begin{equation}\label{Hamiltonian DEs}
\{L\otimes_{,}L\}=[r, L\otimes I]-[r^{T},I\otimes
L]=[r,L_{1}+L_{2}], \end{equation} where
\begin{equation}\label{Hamiltonian DEs}
r=\sum_{\mu,\nu=1}^{N}r^{\mu\nu}X_{\mu}\otimes X_{\nu},\quad
r^{T}=\sum_{\mu,\nu=1}^{N}r^{\nu\mu}X_{\mu}\otimes X_{\nu},\quad
L_{1}=L\otimes I, \quad L_{2}=I\otimes L.
\end{equation}
Finally we have
\begin{equation}\label{Hamiltonian DEs}
\{L\otimes_{,}L\}=(r^{\tau\nu}C^{\mu}_{\tau\lambda}L^{\lambda}-r^{\tau\mu}C^{\nu}_{\tau\lambda}L^{\lambda})X_{\mu}\otimes
X_{\nu}. \end{equation} In this notes we use the following form of
the r-matrix
\begin{equation}\label{Hamiltonian DEs}
r=\sum_{i=1}^{N}E_{ii}\otimes
E_{ii}+2\sum_{i,j=1(i<j)}^{N}E_{ij}\otimes E_{ji}. \quad
\end{equation}

\section{Generalized TL}

In this paper we deal with the system of nonlinear
differential-difference equations
 \begin{equation}
\begin{array}{lcl} \dot p_k& = &
2(a_{k-1}b_{k-1}-a_{k}b_{k})+2uv(\delta_{k 1}-\delta_{k,{-1}}),\\
\dot a_{k}& = & (p_{k}-p_{k+1})a_{k}+2v(b_{k-1}\delta_{k,0}- b_{k+1}\delta_{k,{-1}}),\\
\dot b_{k}& = & (p_{k}-p_{k+1})b_{k}+2u(a_{k-1}\delta_{k,0}- a_{k+1}\delta_{k,{-1}}),\\
\dot u& = & (p_{2}-p_{4})u,\\
\dot v& = & (p_{2}-p_{4})v.
\end{array}
\end{equation}
It transforms to the ordinary TL as $u=v=0$ since we call it as
the generalized TL (1DGTL). The system (3.1) is integrable as it
can be written in Lax form as (2.11) with
\begin{equation}
 L= \begin{pmatrix}  p_{-N} &  a_{-N} &0&\cdots&\cdots&\cdots& \cdots &   \cdots & 0 \cr
                   b_{-N} & \ddots &  \ddots& \ddots &&    & &&\vdots \cr
                    0 & \ddots& \ddots & \ddots &0 & &&&\vdots \cr
                    \vdots& \ddots& \ddots & p_{-1}&a_{-1}&v& &   &\vdots \cr
                   \vdots & & 0 & b_{-1}&p_{0}&a_{0}&0 &  &\vdots \cr
                   \vdots & & & u&b_{0}&p_{1}&\ddots & \ddots  & \vdots\cr
                   \vdots & &  &  & 0&\ddots& \ddots&\ddots& 0 \cr
                   \vdots & &  &  & & \ddots &\ddots&\ddots&a_{N-1}\cr
                   0 & \cdots & \cdots & \cdots & \cdots &\cdots &0&b_{N-1} & p_N     \end{pmatrix},\label{Hamiltonian}
\end{equation}
\begin{equation}
 M= \begin{pmatrix}  0 &  a_{-N} &0&\cdots&\cdots&\cdots& \cdots &   \cdots & 0 \cr
                   -b_{-N} & \ddots& \ddots&  \ddots&&&    & &\vdots\cr
                    0 & \ddots& \ddots & \ddots &0 & &&&\vdots \cr
                    \vdots& \ddots& \ddots & 0&a_{-1}&v& &   &\vdots \cr
                   \vdots & & 0 & -b_{-1}&0&a_{0}&0 &  &\vdots \cr
                   \vdots & & & -u&-b_{0}&0&\ddots & \ddots  & \vdots\cr
                   \vdots & &  &  & 0&\ddots& \ddots&\ddots& 0 \cr
                   \vdots & &  &  & & \ddots &\ddots&\ddots&a_{N-1}\cr
                   0 & \cdots & \cdots & \cdots & \cdots &\cdots &0&-b_{N-1} & 0     
                   \end{pmatrix}.
\end{equation}
There are take place  a  Lie-Poisson brackets (2.13) between
elements of $L$  which are given by
\begin{equation}
\begin{array}{lcl}
\{p_i, a_i \}& =&a_i,  \\
\{p_{i+1}, a_{i} \} &=&- a_i,\\
\{p_i, b_i \}& =&b_i  \\
\{p_{i+1},  b_{i} \} &=& -b_i,\\
\{p_{-1}, u \}_M& =&u,  \\
\{p_{1}, u \}_M &=&-u,   \\
\{p_{-1}, v \}_M& =&v,  \\
\{p_{1}, v \}_M &=&-v,\\
\{a_{-1}, a_0 \}_M &=&2v,\\
\{b_{-1},b_0 \}_M &=&2u.
\end{array}
\label{Hamiltonian}
\end{equation} All other brackets are zero.
  We   denote this bracket by $\pi_1$. The functions
\begin{equation}
H_{i}=\frac{1}{i}trL^{i}
 \label{Hamiltonian}
\end{equation}
are independent invariants in involution that is
\begin{equation}\{  H_i, H_j
\}=0. \label{Hamiltonian}
\end{equation} The expressions for $H_i$ are, for example,
\begin{equation}
\begin{array}{lcl}
H_1=\sum_{k=-N}^Np_k,  \\
H_2=\frac{1}{2}\sum_{k=-N}^{N}p_{k}^{2}+\sum_{k=-(N-1)}^{N-1}a_{k}b_k
+uv,\\
H_3=\frac{1}{3}\sum_{k=-N}^{N}p_{k}^{3}+\sum_{k=1}^{N}[a_{1}b_1p_1+(a_1b_1+a_2b_2+uv)p_2+
(a_2b_2+a_3b_3)p_3+\\(a_3b_3+a_4b_4+uv)p_4+a_4b_4p_5+a_2a_3u+b_2b_3v],
\end{array}
 \label{Hamiltonian}
\end{equation}
and so on. The invariant $H_1$ is the only Casimir. The
Hamiltonian in this bracket is
 $H_2 = { 1 \over 2}\  { \rm tr}\  L^2$.

As for the usual TL in our case we can introduce the quadratic
Toda brackets which  appears in conjunction with isospectral
deformations of Jacobi matrices. It is a Poisson bracket in which
the Hamiltonian vector field generated by $H_1$ is the same as the
Hamiltonian vector field generated by $H_2$ with respect to the
$\pi_1$ bracket.  We will denote this Poisson bracket by $\pi_2$.
The bracket $\pi_2$ is easily defined by taking  the Lie
derivative of $\pi_1$  in the direction of  suitable master
symmetry. This bracket has ${\rm det} \, L$ as Casimir and $H_1
={\rm tr}\, L$
 is the Hamiltonian. The eigenvalues of $L$ are still in involution.
Furthermore, $\pi_2$ is compatible with $\pi_1$.
 We also have
\begin{equation}
 \pi_2 d H_i = \pi_1   d H_{i+1}.
  \label{Hamiltonian}
\end{equation}
Note that both brackets $\pi_1$ and $\pi_2$  transforms to the
corresponding brackets of the usual TL for the case $u=v=0$.
\section{Case N=3}
%\subsubsection{$b_k=a_k$}
Let us now we consider in more detail the case $N=3$. For this case  the corresponding 
equations of the 1DGTL take the forms
\begin{equation}
\begin{array}{lcl} \dot p_1& = &
-2(a_{1}^2+u^2),\\
\dot p_2& = &
2(a_{1}^2-a_{2}^2),\\
\dot p_3& = &
2(a_{2}^2+u^2),\\
\dot a_{1}& = & a_1(p_{1}-p_{2})-2ua_2,\\
\dot a_{2}& = & a_2(p_{2}-p_{3})+2ua_1,\\
\dot u& = & (p_{1}-p_{3})u.
\end{array}\label{Hamiltonian}
\end{equation}
This system can be written in Lax form as
\begin{equation}
\dot L=[L,M], \label{Hamiltonian}
\end{equation}
where
\begin{equation}
 L= \begin{pmatrix}  p_1 &  a_1 & u \cr
                   a_1 & p_2 & a_2  \cr
                   u & a_2 & p_3      \end{pmatrix},\quad
 M=  \begin{pmatrix}  0 &  a_1 & u \cr
                   -a_1 & 0 & a_2  \cr
                   -u & -a_2 & 0      \end{pmatrix}.\label{Hamiltonian}
\end{equation}
 There exists a
Lie-Poisson bracket given by the formula
\begin{equation}
\begin{array}{lcl}
\{p_i, a_i \}& =&a_i,  \\
\{p_{i+1}, a_{i} \} &=&- a_i,\\
\{p_1, u \}& =&u,  \\
\{p_3, u \} &=&-u,   \\
\{a_1, a_2 \} &=&2u.
\end{array}
\label{Hamiltonian}
\end{equation} All other brackets are zero.
  We   denote this bracket by $\pi_1$. The functions
\begin{equation}
H_{i}=\frac{1}{i}trL^{i} \label{Hamiltonian}
\end{equation}
are independent invariants in involution that is
\begin{equation} \{  H_i, H_j
\}=0.$$ The expressions for $H_i$ are, for example, $$
\begin{array}{lcl}
H_1=\sum_{k=1}^3p_k,  \\
H_2=\frac{1}{2}\sum_{k=1}^{3}p_{k}^{2}+\sum_{k=1}^{2}a_{k}^2+u
^2,\\
H_3=\frac{1}{3}\sum_{k=1}^{3}p_{k}^{3}+\sum_{k=1}^{3}[a_{1}^2p_1+(a_1^2+a_2^2+u^2)p_2+
(a_2^2+a_3^2)p_3+\\(a_3^2+a_4^2+u^2)p_4+a_4^2p_5+2a_2a_3u]
\end{array}
\label{Hamiltonian}
\end{equation}
and so on. The Hamiltonian in this bracket is
 $H_2 = { 1 \over 2}\  { \rm tr}\  L^2.$ The Casimirs of the system are
\begin{equation}
\begin{array}{lcl} C_1& = &
H_1=p_{1}+p_2+p_3,\\
C_2& = &\frac{a_{1}a_{2}}{u}-2p_2,\\
C_3& = & \frac{v}{u}=d_3.
\end{array}\label{Hamiltonian}
\end{equation}  Note that $$\dot C_k=0.$$

\subsection{The  $q_k, p_k$ coordinates }

Let us rewrite the system (4.1) in terms of the  coordinates  $q_{k},
\quad p_{k}=\dot q_k, \quad k=1,2,3$. Then from (4.1) we have
\begin{equation}
\begin{array}{lcl} \ddot q_1& = &
-2(a_{1}^2+u^2),\\
\ddot q_2& = &
-2(a_{2}^2-a_{1}^2),\\
\ddot q_3& = &
2(a_{2}^2+u^2),\\
\dot a_{1}& = & a_1(p_{1}-p_{2})-2ua_2,\\
\dot a_{2}& = & a_2(p_{2}-p_{3})+2ua_1,\\
\dot u& = & p_{13}u.
\end{array}\label{Hamiltonian}
\end{equation}
Hence we get
\begin{equation}
u=u_0+e^{q_{13}},\quad q_{ij}=q_i-q_j,\quad p_{ij}=p_i-p_j, ,\quad
u_0=const(t).\label{Hamiltonian}
\end{equation}
Let $$a_1=e^{q_{12}}p_4, \quad a_2=e^{q_{23}}q_4.$$ Then
\begin{equation}
 L= \begin{pmatrix} \ p_1 &  e^{q_{12}}p_4 & e^{q_{13}} \cr
                   d_1e^{q_{12}}p_5 & p_2 & e^{q_{23}}q_4  \cr
                   d_1d_2e^{q_{13}} & d_2e^{q_{23}}q_5 & p_3      \end{pmatrix},\label{Hamiltonian}
\end{equation}
and the equations of motion become
\begin{equation}
\begin{array}{lcl} \ddot q_1& = &
-2(a_{1}^2+u^2),\\
\ddot q_2& = &
-2(a_{2}^2-a_{1}^2),\\
\ddot q_3& = &
2(a_{2}^2+u^2),\\
\dot p_{4}& = & -2e^{q_{13}+q_{23}-q_{12}}q_{4}=-2e^{2q_{23}}q_{4},\\
\dot q_{4}& = & 2e^{q_{13}-q_{23}+q_{12}}p_{4}=2e^{2q_{12}}p_{4},\\
\dot u& = & p_{13}u.
\end{array}\label{Hamiltonian}
\end{equation}
So we see that $p_{4}=\frac{1}{2}e^{2q_{21}}\dot q_4\neq \dot
q_{4}$. Finally we have
\begin{equation}
\begin{array}{lcl} \ddot q_1& = &
-2(e^{2q_{12}}p_{4}^2+e^{2q_{13}}),\\
\ddot q_2& = &
-2(e^{2q_{23}}q_4^2-e^{2q_{12}}p_4^2),\\
\ddot q_3& = &
2(e^{2q_{23}}q_4^2+e^{2q_{13}})\\
\ddot q_{4}& = & 4(p_{12}p_{4}e^{2q_{12}}-q_{4}e^{2q_{13}}),\\
\dot u& = & p_{13}u
\end{array}\label{Hamiltonian}
\end{equation}
or
\begin{equation}
\begin{array}{lcl} \ddot q_1& = &
-2(p_{4}^2e^{2q_{12}}+e^{2q_{13}}),\\
\ddot q_2& = &
-2(q_4^2e^{2q_{23}}-p_4^2e^{2q_{12}}),\\
\ddot q_{4}& = & 4(p_{12}p_{4}e^{2q_{12}}-q_{4}e^{2q_{13}}).
\end{array}\label{Hamiltonian}
\end{equation}
Here
\begin{equation} C_4=p_4q_4+\alpha e^{p_4q_4}.
\label{Hamiltonian}
\end{equation}
\subsection{The  ($P_k, Q_k$) coordinates }

Let us consider a new representation for the Lax matrix $L$
as
\begin{equation}
 L= \begin{pmatrix} P_1 &  e^{Q_{1}}P_3 & e^{Q_{1}+Q_{2}} \cr
                   d_1e^{Q_{1}}P_4 & P_2-P_1 & e^{Q_{2}}Q_3  \cr
                   d_1d_2e^{Q_{1}+Q_{2}} & d_2e^{Q_{2}}Q_4 & -P_2      \end{pmatrix},\label{Hamiltonian}
\end{equation}where $\{P_i,Q_j\}=\delta_{ij}$. In this case the equations of motion takes the form
\begin{equation}
\begin{array}{lcl} \dot P_1& = &Q_1,\\
\dot P_2& = & Q_2,\\
\dot P_3& = & Q_3,\\
\dot Q_{1}& = & a_1(p_{1}-p_{2})-2ua_2,\\
\dot Q_{2}& = & a_2(p_{2}-p_{3})+2ua_1,\\
\dot Q_{2}& = & a_2(p_{2}-p_{3})+2ua_1,\\
\dot u& = & (P_{1}-P_{3})u.
\end{array}\label{Hamiltonian}
\end{equation}

\section{Solutions} 

To find solutions we first introduce a new
variables as
\begin{equation}
c_k(t)=p_{k}(\frac{1}{2}t),\quad
d_{k+1}(t)=a^2_{k}(\frac{1}{2}t),\quad w=u^2(\frac{1}{2}t).
\label{Hamiltonian}
\end{equation} Then
the system (4.1) becomes
\begin{equation}
\begin{array}{lcl} \dot c_1& = &
-(d_{2}+w),\\
\dot c_2& = &
d_2-d_{3},\\
\dot c_3& = &
d_3+w\\
\dot d_{2}& = &c_{12} d_2-2\sqrt{wd_3},\\
\dot d_{3}& = & c_{23}d_3+2\sqrt{wd_2},\\
\dot w& = & c_{13}w.
\end{array}
\end{equation}
Note that the GTL equation (4.18) is the isospectral
($\lambda_t=0$) and is the  compatibility condition of the following spectral
problems
\begin{equation}
\begin{array}{lcl} \dot \phi_n =c_{n-1}\phi_n+\phi_{n-1},\\
d_n\phi_{n+1}+c_{n-1}\phi_n+\phi_{n-1}=\lambda \phi_n,
\end{array}\label{Hamiltonian}
\end{equation}
where $\lambda$ is a spectral parameter. This means that  it has the Lax
representation $L_t=[M,L]$, where the Lax pair is defined by
\begin{equation}
\begin{array}{lcl} L_{nm}& = &
d_{n}\delta_{n+1,m}+c_{n-1}\delta_{nm}+\delta_{n-1,m},\\
M_{nm}& = & \delta_{n+1,m}+c_{n-1}\delta_{nm}.
\end{array}\label{Hamiltonian}
\end{equation}
In the matrix form these matrices have the form
\begin{equation}
 L= \begin{pmatrix}  c_0 &  d_1 & w \cr
                   1 & c_1&d_2  \cr
                   0 & 1 & c_2      \end{pmatrix}, \quad M= \begin{pmatrix}  c_0 &  1 & 0 \cr
                   0& c_1&1  \cr
                   0 & 0 & c_2      \end{pmatrix},\label{Hamiltonian}
\end{equation}
If we introduce the following $\tau$-function form of the
dependent variable
\begin{equation}
d_{k}=1+(\ln\tau_{k})_{tt}=(\ln
\tau_ke^{\frac{1}{2}t^2})_{tt},\quad w=-1-(\ln f)_{tt}=-(\ln
fe^{\frac{1}{2}t^2})_{tt}, \label{Hamiltonian}
\end{equation}then we have
\begin{equation}
\begin{array}{lcl}  c_1& = &
I_1-(\ln{\frac{\tau_2}{f}})_t,\\
c_2& = &
I_2+(\ln{\frac{\tau_{2}}{\tau_3}})_t,\\
c_3& = & I_3+(\ln{\frac{\tau_3}{f}})_t
\end{array}\label{Hamiltonian}
\end{equation}
and hence
\begin{equation}
\begin{array}{lcl}  c_{12}& = &
I_{12}-(\ln{\frac{\tau_2^2}{\tau_3f}})_t,\\
c_{23}& = &
I_{23}+(\ln{\frac{\tau_{2}f}{\tau_3^2}})_t,\\
c_{13}& = & I_{13}-(\ln{\frac{\tau_2\tau_3}{f^2}})_t.
\end{array}\label{Hamiltonian}
\end{equation}
To define the unkown functions we have the system
\begin{equation}
\begin{array}{lcl}
(d_{2}+d_3)_t&=&c_{12}d_2+c_{23}d_3,\\
(\dot d_{3}-c_{23}d_3)^2&=&4wd_2,\\
\dot w-c_{13}w& = & 0.
\end{array}\label{Hamiltonian}
\end{equation}
Hence and from (5.6) and (5.8) we get
\begin{equation}
\begin{array}{lcl}
(\ln{\tau_2\tau_3})_{ttt}-[I_{12}-(\ln{\frac{\tau_2^2}{\tau_3f}})_t](\ln
\tau_2e^{\frac{1}{2}t^2})_{tt}-[I_{23}+
(\ln{\frac{\tau_{2}f}{\tau_3^2}})_t](\ln
\tau_3e^{\frac{1}{2}t^2})_{tt} &=&0, \\
\{(\ln\tau_3)_{ttt}-[I_{23}+(\ln{\frac{\tau_{2}f}{\tau_3^2}})_t](\ln
e^{\frac{1}{2}t^2}\tau_3)_{tt}\}^2 +4(\ln
\tau_2e^{\frac{1}{2}t^2})_{tt}(\ln
fe^{\frac{1}{2}t^2})_{tt}&=&0,\\
\ddot ff-\dot
f^2+f^2+\frac{f^4}{\tau_2\tau_3}e^{I_{130}+I_{13t}}&=&0.

\end{array}\label{Hamiltonian}
\end{equation}
For simplicity we set $I_{ij}=0$. Then the system (5.10) becomes
\begin{equation}
\begin{array}{lcl}
(\ln{\tau_2\tau_3})_{ttt}+(\ln{\frac{\tau_2^2}{\tau_3f}})_t(\ln
\tau_2e^{\frac{1}{2}t^2})_{tt}-
(\ln{\frac{\tau_{2}f}{\tau_3^2}})_t(\ln
\tau_3e^{\frac{1}{2}t^2})_{tt} &=&0 \\

[(\ln\tau_3)_{ttt}-(\ln\frac{\tau_{2}f}{\tau_3^2})_t(\ln
e^{\frac{1}{2}t^2}\tau_3)_{tt}]^2 +4(\ln
\tau_ke^{\frac{1}{2}t^2})_{tt}(\ln fe^{\frac{1}{2}t^2})_{tt}&=&0\\
\ddot ff-\dot f^2+f^2+\frac{f^4}{\tau_2\tau_3}e^{I_{130}}&=&0.

\end{array}\label{Hamiltonian}
\end{equation}
We expand the functions $\tau_n, f$ in a formal power series in an
arbitrary parameter $\epsilon$ as
\begin{equation}
\begin{array}{lcl}
\tau_n=\sum_{k=0}\epsilon^k\tau_{n}^{(k)},\quad
f=\sum_{k=0}\epsilon^kf^{(k)}.\end{array}\label{Hamiltonian}
\end{equation}
Expanding the l.h.s. of (5.11) in $\epsilon$ and equating
corresponding coefficients, we get the resulting equations for the
functions $\tau_{n}^{(k)}, f^{(k)}$. Solving these equations we
can construct an exast solutions of the underlying set of
equations for the 1DGTL.

\section{The 1DGTL hierarchy}

First let us recall the main formulas of the usual TL hierarchy.
The corresponding  hierarchy is defined by
\begin{equation}
\frac{\partial L}{\partial t_{k}}=[L,B_{k}], \quad
B_{k}=(L^{k})_{-},\quad k=1,2,3, ... \, .
\end{equation}
The $\tau$-functions of the TL hierarchy obey the following
equations
\begin{equation}
[D_{k}-h_{k}({\bf D})]\tau_{n+1}\cdot \tau_{n}=0, \quad k=2,3,4,
... \,.
\end{equation}
Here
\begin{equation}
e^{\sum^{\infty}_{k=1}\frac{1}{k}D_kz^k}=\sum^{\infty}_{n=0}h_{n}({\bf
D})z^{n}, \quad {\bf D}=(D_1, \frac{1}{2}D_2, \frac{1}{3}D_3,
\frac{1}{4}D_4, ... ).\label{Hamiltonian}
\end{equation}
It is interesting to note that the nonlinear Schrodinger equation
(NLSE) is the second member of the TL hierarchy. In fact from
(6.2) as $k=2$ and from (2.6) we get the following set of
equations (see for example \cite{Kodama-Pierce})
\begin{equation}
\begin{array}{lcl}
(D_2-D^2_1)\tau_{n+1}\cdot\tau_{n}=0,\\
D^2_1\tau_{n}\cdot\tau_{n}=2\tau_{n+1}\cdot\tau_{n-1}.\end{array}\label{Hamiltonian}
\end{equation}
This set is equivalent to the NLSE \cite{Kodama-Pierce}
\begin{equation}
i\phi_{t_2}+\phi_{t_1t_1}+2\phi^2\bar\phi=0,\label{Hamiltonian}
\end{equation}
where
\begin{equation}
\phi=\tau_{n+1}\tau^{-1}_{n},\quad
\bar\phi=\tau_{n-1}\tau^{-1}_{n}, \quad t_2\longrightarrow
it_2.\label{Hamiltonian}
\end{equation}
The set (6.4) is the compatibility condition of the following set
of linear equations
\begin{equation}
 \psi_{t_1}=\begin{pmatrix}  0 & \tau_{n-1}\tau^{-1}_{n}  \cr
                  \tau_{n+1}\tau^{-1}_{n} & 0
                  \end{pmatrix}\psi,\quad
 \psi_{t_2}=i\begin{pmatrix}  \tau_{n-1}\tau_{n+1}\tau^{-2}_{n} & -(\tau_{n-1}\tau^{-1}_{n})_{t_1}  \cr
                  (\tau_{n+1}\tau^{-1}_{n})_{t_1} & -\tau_{n-1}\tau_{n+1}\tau^{-2}_{n}      \end{pmatrix}\psi. \label{Hamiltonian}
\end{equation}
Note that in this case the matrix $S=\psi^{-1}\sigma_3\psi$ obeys
the Heisenberg ferromagnetic equation
\begin{equation}
2iS_{t_2}=[S,S_{t_1t_1}].\label{Hamiltonian}
\end{equation}
Finally note that the 1DGTL hierarchy has the same form as (6.1)
but content the additional two equations.
\section{Conclusion }
In the present Letter we considered one of integrable
generalizations of TL. The corresponding Lax representation is
presented. For the particular case $N=3$ the bilinear form
($\tau$-function form) is found that allows to construct exast
solutions of the studied generalized Toda equation.

\section*{Acknowledgments}
One of the authors (R.M.) thanks   N. Reshetikhin for the statement of the problem anf for  very helpful discussions.  R.M. also would like to thank  N.Reshetikhin and Department of Mathematics, UC, Berkeley  for their hospitality during his visit (January-May, 2009).

\end{document}